\documentclass[a4paper,11pt]{article}
\usepackage{pos}
\usepackage{xspace}
\usepackage{float}
\usepackage{amsmath}
\newcommand{\pt}           {\ensuremath{p_{\rm T}}\xspace}

\newcommand{\RAA}          {\ensuremath{R_{\rm AA}}\xspace}

\newcommand{\vtwo}         {\ensuremath{v_{\rm 2}}\xspace}

\newcommand{\fnonprompt}   {\ensuremath{f_\mathrm{np}}\xspace}

\newcommand{\GeVc}         {\ensuremath{\mathrm{GeV}/c}\xspace}

\newcommand{\DzerotoKpi}        {\ensuremath{{\rm D}^0 \to {\rm K}^-\pi^+}\xspace}
\newcommand{\DplustoKpipi}      {\ensuremath{{\rm D}^+\to {\rm K}^-\pi^+\pi^+}\xspace}

\newcommand{\DstophipitoKKpi}   {\ensuremath{{\rm D_s^{+}\to \upphi\pi^+\to K^-K^+\pi^+}}\xspace}

\newcommand{\Dzero}             {\ensuremath{\mathrm{D^0}}\xspace}

\newcommand{\Dplus}             {\ensuremath{\mathrm{D^+}}\xspace}

\newcommand{\Ds}                {\ensuremath{\mathrm{D_s^+}}\xspace}
\newcommand{\Lc}                {\ensuremath{\Lambda_\mathrm{c}^+}\xspace}

\newcommand{\LctopKpi}          {\ensuremath{\Lambda_\mathrm{c}^+\to\mathrm{pK^-\pi^+}}\xspace}

\newcommand{\Jpsi}                {\ensuremath{\mathrm{J}/\uppsi}\xspace}

\title{Investigating charm-quark dynamics in the QGP via the charm-hadron elliptic flow in Pb–Pb collisions with ALICE}
\ShortTitle{Charm-hadron elliptic flow in Pb–Pb collisions with ALICE}

\author*[a,b]{Marcello Di Costanzo \onbehalf{on behalf of the ALICE Collaboration}}

\affiliation[a]{Polytechnic University,\\
  Corso Castelfidardo 39, Turin, Italy}

\affiliation[b]{INFN Turin,\\
  Via P. Giuria 1, Turin, Italy}

\emailAdd{marcello.di.costanzo@cern.ch}

\abstract{
  In these proceedings, the elliptic flow ($\vtwo$) measurement of prompt and non-prompt charm hadrons—originating respectively from the fragmentation of a charm quark and from the decay of hadrons with beauty-quark content—in Pb–Pb collisions at $\sqrt{s_{\mathrm{NN}}} = 5.36$ TeV, using the latest data collected during LHC Run 3 by the ALICE detector, is presented.
  The analysis is performed at midrapidity ($|y| < 0.8$) and hadronic decay channels are used to reconstruct the signal candidates.
  The $\vtwo$ coefficient is measured via the Scalar Product (SP) technique.
  The prompt $\vtwo$ in semicentral collisions is reported for the $\Dzero$, $\Dplus$, $\Ds$ mesons and, for the first time, for the $\Lc$ baryon.
  These results achieve unprecedented precision and low-$\pt$ reach for the D mesons and highlight the first observation of the splitting of baryon and meson elliptic flow at intermediate $\pt$ in the charm sector.
  The $\vtwo$ is also measured for prompt $\Dzero$, $\Dplus$ and $\Ds$ in peripheral collisions, offering new insights into the features of the fireball generated in ultrarelativistic heavy-ion collisions.
  Lower $\vtwo$ values are observed for non-prompt $\Dzero$ and $\Lc$ hadrons in semicentral collisions, as expected from the larger beauty-quark mass.
}

\FullConference{The European Physical Society Conference on High Energy Physics (EPS-HEP2025)\\
7-11 July 2025\\
Marseille, France\\}

\begin{document}
\maketitle
\section{Introduction}
The quark--gluon plasma (QGP) is a color deconfined state of matter predicted by Lattice Quantum Chromodynamics (QCD) calculations to exist at low baryochemical potential and temperatures above 155 MeV.
The study of QCD matter in these extreme conditions relates to the fundamental properties of QCD and to the evolution of the early universe.
The above described conditions can be attained in ultrarelativistic heavy-ion collisions.
The generated fireball, featuring partonic degrees of freedom, expands and cools down until hadronization occurs at $T \approx 155$ MeV~\cite{Ratti:2018ksb}, leaving the system in a dense hadronic gas state.
Interactions among hadrons can still occur but, as the system continues to expand and dilute, it reaches the kinetic freeze-out.
In this scenario, heavy-flavor hadrons are unique probes to investigate the QGP properties, as heavy quarks are produced in hard scatterings occurring before the QGP formation, thus experiencing the full medium evolution.
Theoretical descriptions of QGP and hadronization are hindered by the non-perturbative nature of QCD in the relevant regime, making phenomenological transport models the main tool to describe observables.
In this approach, heavy-quark interactions with the QGP are encoded in model parameters, primarily the spatial diffusion coefficient $D_S$, which can be constrained by measurements of heavy-flavor hadron production.\\

The interactions of the quarks in the QGP medium can be studied through the elliptic flow observable.
In non-central ultrarelativistic heavy-ion collisions, the almond-shaped overlap region of the colliding nuclei is understood to be the origin of elliptic flow~\cite{Liu:2012ax}.
Due to the initial geometrical anisotropy, the generated fireball expands under non-isotropic pressure gradients that produce a modulation of the particle yields with respect to the azimuthal angle $\varphi$ relative to the reaction plane (RP).
The latter is identified as the plane containing the impact parameter of the collision and the beam axis, and its azimuthal angle is indicated by $\Psi_{\text{RP}}$.
The particle yield modulation with respect to $\varphi-\Psi_{\text{RP}}$, with $\varphi$ indicating the azimuthal angle of the emitted particle, can be described via a Fourier series expansion
\begin{equation}
    \frac{\text{d}N}{\text{d}\varphi} = \frac{N_0}{2\pi} \bigg( 1 + 2 \sum_{n=1}^{\infty} v_{n} (\cos[n(\varphi - \Psi_{\text{RP}})]) \bigg), \qquad v_n = \langle \cos[n(\varphi - \Psi_{\text{RP}})] \rangle,
\end{equation}
where $\vtwo$ represents the elliptic flow coefficient, sensitive to the initial geometry of the collision and to the interactions of the charm and beauty quarks in the QGP medium.
It can be argued that the hadronization mechanisms of charm hadrons have direct implication on their $\vtwo$ coefficient.
Indeed, when the hadron is produced in the fragmentation of a charm quark, its $\vtwo$ will solely feature the contribution from the interactions of the charm quark in the QGP medium.
Instead, when the hadron is produced via coalescence of a heavy quark with light quarks from the QGP~\cite{Zhao:2020jqu}, its $\vtwo$ will also include the contribution from the interactions of the captured light quarks.
Lastly, when the hadron, referred to as non-prompt in this case, is produced in the decay of a hadron with beauty-quark content, its $\vtwo$ reflects the interactions of the beauty quark in the QGP medium, with a slight smearing due to the decay kinematics.
The elastic and inelastic rescattering processes occurring in the hadron gas phase before the freeze-out can further contribute to the measured $\vtwo$.

\section{Analysis procedure}
\begin{figure}[bt]
    \centering
    \includegraphics[width=0.75\textwidth]{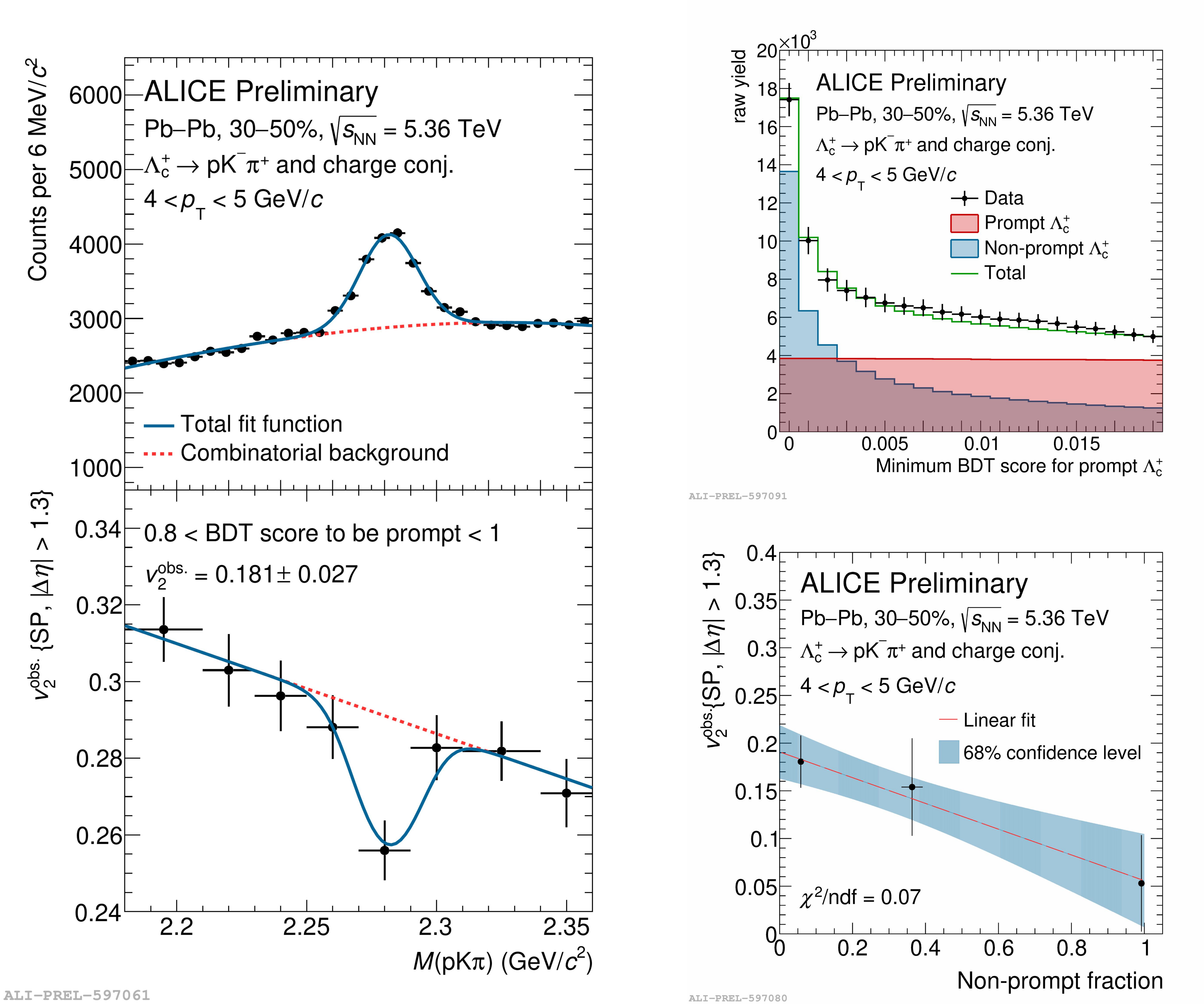}
    \caption{Left: simultaneous fit of the invariant mass distribution and $\vtwo$ as a function of the mass. Top right: $\Lc$ raw yields as a function of the BDT prompt score selection with templates from prompt and non-prompt contributions obtained from Monte Carlo simulations. Bottom right: linear fit to the observed inclusive $\vtwo$ values for different values of non-prompt fraction.}
    \label{fig:analysis_technique}
\end{figure}

In this proceedings, we present the measurements of the $\vtwo$ of prompt and non-prompt $\Dzero$ and $\Lc$ hadrons, and of prompt $\Dplus$ and $\Ds$ mesons, in Pb–Pb collisions at $\sqrt{s_{\mathrm{NN}}} = 5.36$ TeV for several centrality classes.
For the analysis, the entire data sample recorded by ALICE in 2023 is employed, amounting to an integrated luminosity of $\mathcal{L}_{\text{int}} \approx 1.5$ nb$^{-1}$.
The ITS, TPC, and TOF detectors reconstruct charged particles at midrapidity ($|y|<0.8$) and provide particle identification (PID) information, with improved performance compared to LHC Run~2 owing to major upgrades implemented during LHC Long Shutdown~2~\cite{ALICE:2023udb}.
The $\Dzero$, $\Dplus$, $\Ds$ mesons and $\Lc$ baryons are reconstructed in selected hadronic decay channels along with their charge conjugates: $\DzerotoKpi$ (BR $= 3.95\pm0.03\%$), $\DplustoKpipi$ (BR $= 9.38\pm0.16\%$), $\DstophipitoKKpi$ (BR $= 2.22\pm0.06\%$), and $\LctopKpi$ (BR $= 6.28\pm0.32\%$)~\cite{ParticleDataGroup:2024cfk}.
The displaced decay-vertex topologies determined by the lifetime of charm and beauty hadrons, of the order of 100 $\mu$m and 400 $\mu$m respectively, along with the PID information of their decay products are exploited by Machine Learning based multiclass BDT models~\cite{10.1145/2939672.2939785} to separate the background, prompt, and non-prompt contributions.
The Scalar Product (SP) method~\cite{Poskanzer:1998yz} is employed to extract the $\vtwo$ signals and the FT0C, FV0, and TPC detectors are used to estimate the $Q$-vector resolution~\cite{Luzum:2012da}.
Since the $\vtwo$ signal cannot be extracted on a candidate-by-candidate basis, a simultaneous fit to the invariant mass distribution and to the candidate $\vtwo$ in intervals of invariant mass is performed.
An example is reported in the left panel of Figure \ref{fig:analysis_technique}, which shows a clear reduction of the candidate $\vtwo$ around the $\Lc$ signal peak, indicating a distinct elliptic flow for signal compared to the combinatorial background.

Since the prompt and non-prompt charm-hadron $\vtwo$ coefficients differ and exclusive samples of each contribution cannot be obtained, corrections are applied to extract the respective $\vtwo$ values.
For all extracted values, the fraction of non-prompt candidates $\fnonprompt$ is estimated in each $\pt$ interval via a data-driven procedure~\cite{cutvariationdmesprod} in which raw yields from invariant-mass fits and acceptance and efficiency factors from Monte Carlo simulations, distinct for the prompt and non-prompt cases, are evaluated for sequential BDT selections.
As shown in the top right panel of Figure \ref{fig:analysis_technique}, a good description of the observed variation of the raw yields with the prompt BDT score fixes the corrected yield of the two contributions.
Given the large size of the employed data sample, for each $\pt$ interval several independent $\vtwo$ measurements at different $\fnonprompt$ values can be obtained.
As in Ref.~\cite{npd0v2} (bottom right of Figure \ref{fig:analysis_technique}), the values are linearly fitted to extrapolate the fully corrected prompt and non-prompt $\vtwo$ at $\fnonprompt=0$ and $\fnonprompt=1$, respectively.

\section{Results}
\begin{figure}[h]
\centering
\includegraphics[width=0.99\textwidth]{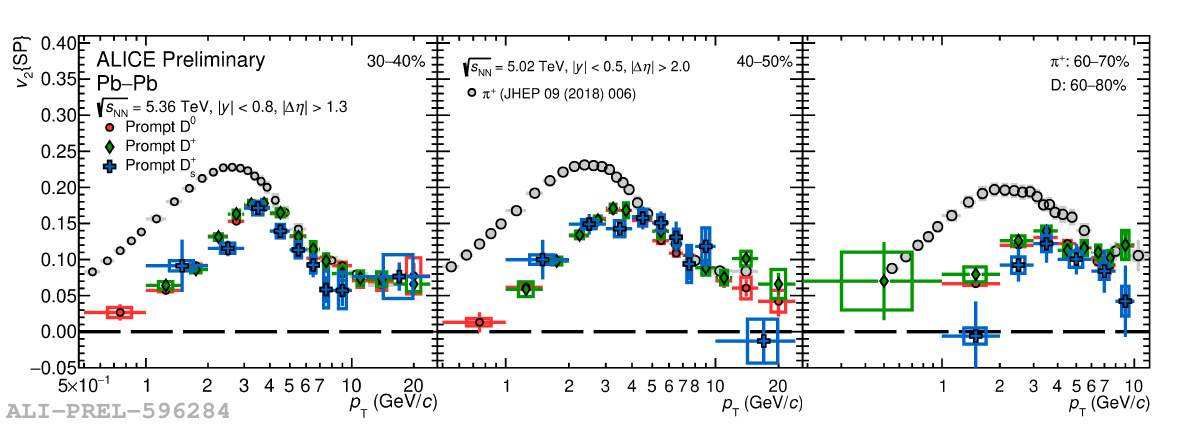}
\caption{Prompt $\Dzero$, $\Dplus$, $\Ds$ $\vtwo$ in Pb–Pb collisions at $\sqrt{s_{\mathrm{NN}}} = 5.36$ TeV for the centrality percentiles $30$–$40\%$, $40$–$50\%$ and $60$–$80\%$. The grey markers indicate the pion $\vtwo$ from Ref.~\cite{pionv2}.}
\label{fig:prompt_dmeson_v2_run3_vscentpass4}
\end{figure}

Figure \ref{fig:prompt_dmeson_v2_run3_vscentpass4} shows the measured prompt $\Dzero$, $\Dplus$, and $\Ds$ $\vtwo$ for the $30$–$40\%$ and $40$–$50\%$ centrality classes and for the peripheral $60$–$80\%$ centrality class, measured for the first time for charm hadrons.
It can be seen that an excellent precision is accomplished for all D meson species over a wide $\pt$ range, with the measurements of non-strange D mesons extending down to 1 $\GeVc$.
Firstly, a remarkable agreement is observed between the $\vtwo$ values of $\Dzero$ and $\Dplus$ mesons.
A reduction of the $\vtwo$ is seen in peripheral events, where the fireball has larger initial geometrical anisotropy but is shorter-lived, which limits both the bulk flow development and the charm-quark thermal equilibration with the medium.
Lastly, it is interesting to compare the D-meson $\vtwo$ curves with the one of pions, reported in grey and taken from a previous ALICE measurement~\cite{pionv2}.
the D-meson $\vtwo$ is similar in magnitude to the pion $\vtwo$, indicating that at least a partial thermal equilibration of the charm quarks in the QGP medium has occurred.
At low $\pt$, the mass ordering predicted by hydrodynamic models of the fireball expansion is clearly observed, with the pion $\vtwo$ being significantly larger than the D-meson $\vtwo$.
At intermediate $\pt$ ($3$$\,<\,$$\pt$$\,<\,$$8$ $\GeVc$), the pion and D-meson curves converge.
This supports coalescence as the dominant production mechanism, since the recombination probability depends on the spatial distribution of light quarks.  
The effect is stronger for baryons than for mesons due to the larger number of constituent quarks, so baryons are expected to exhibit a higher $\vtwo$ where coalescence dominates.

\begin{figure}[h]
\centering
\includegraphics[width=0.415\textwidth]{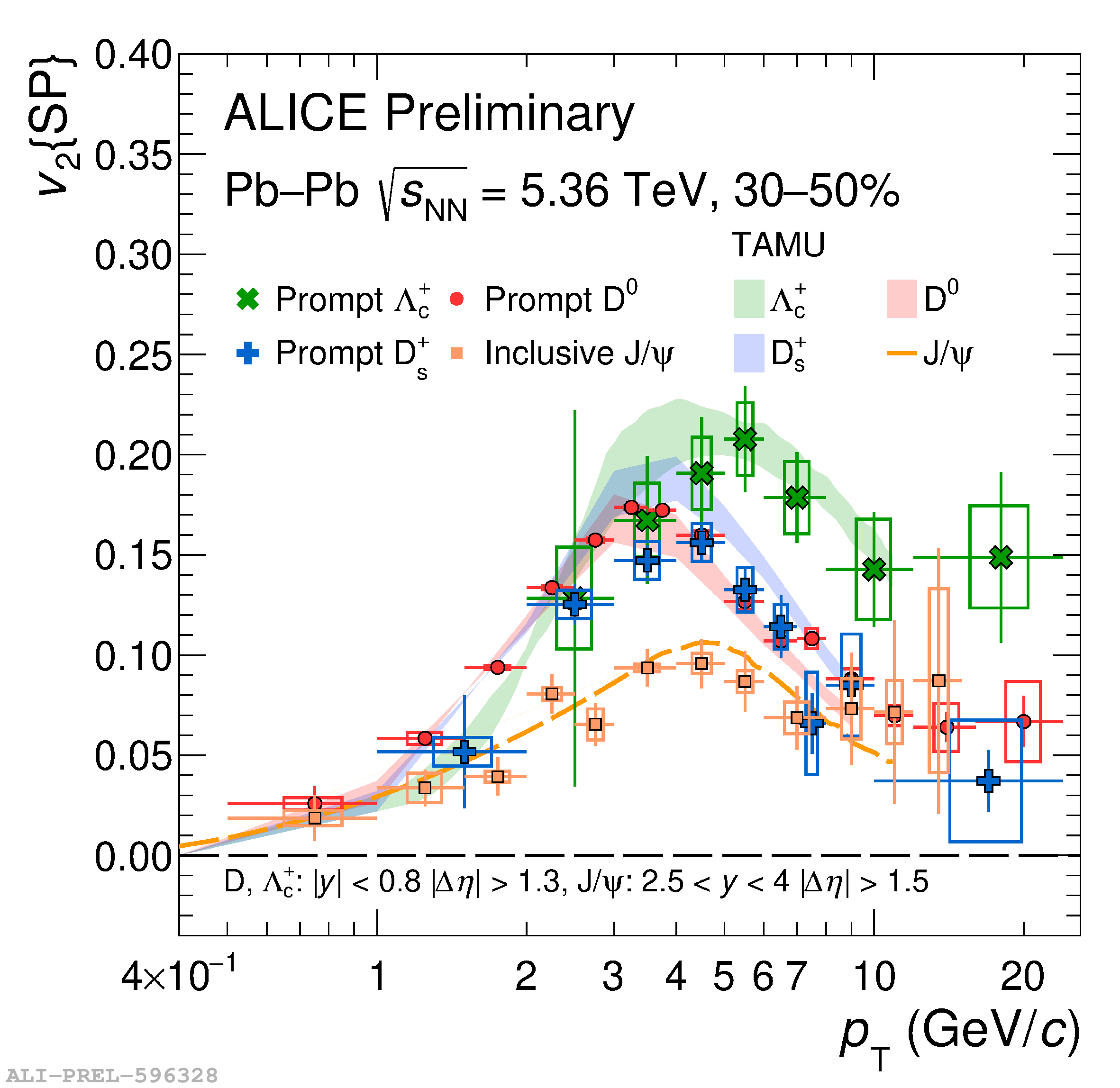}
\includegraphics[width=0.575\textwidth]{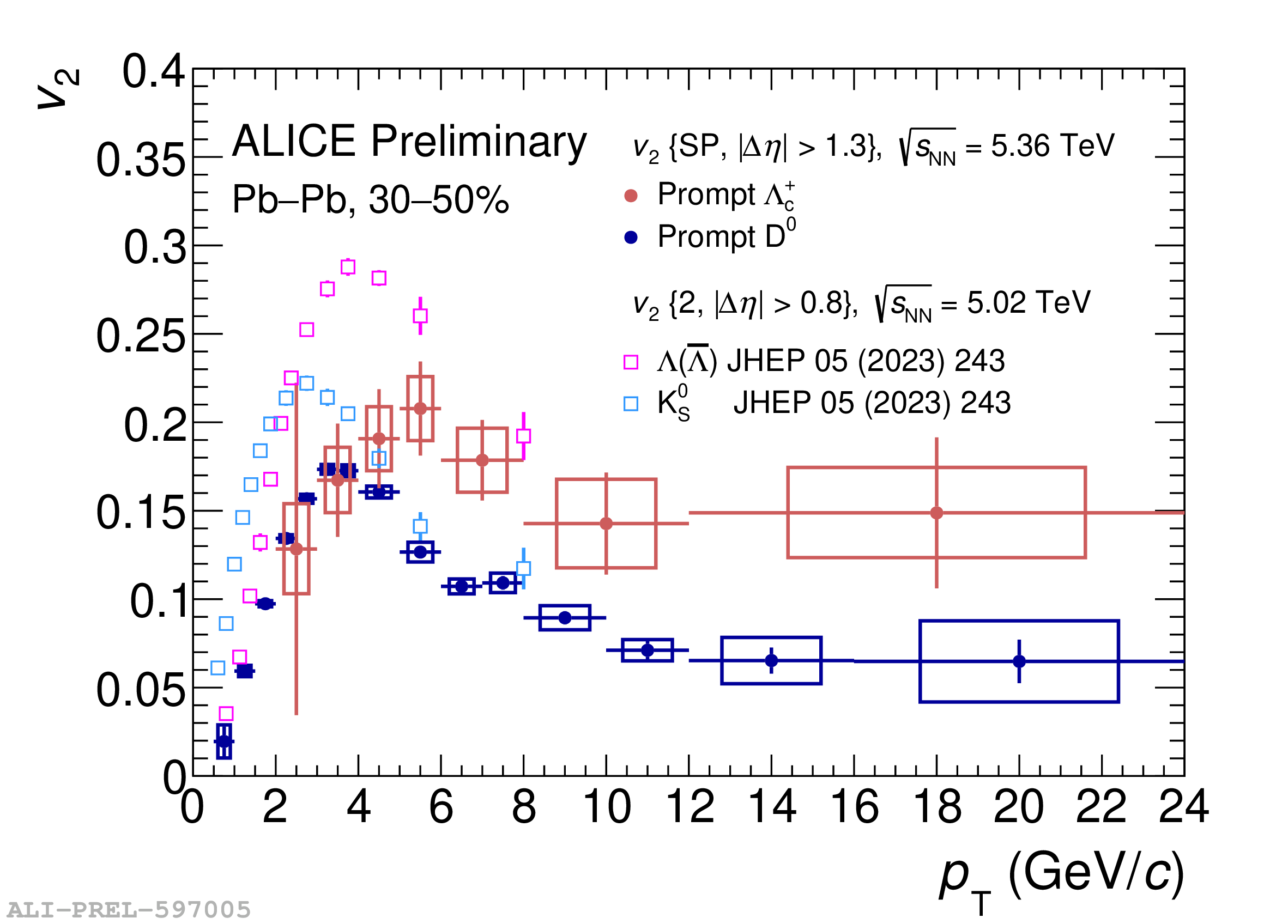}
\caption{Left: elliptic flow of prompt $\Dzero$, $\Ds$ and $\Lc$, and inclusive $\Jpsi$ (from Ref.~\cite{jpsi_v2}) in semicentral Pb–Pb collisions at $\sqrt{s_{\mathrm{NN}}} = 5.36$ TeV superimposed with TAMU model~\cite{He:2019vgs} predictions. Right: superposition of prompt $\Dzero$ and $\Lc$ $\vtwo$ measurements with $\Lambda$ and $\text{K}^0_\text{S}$ $\vtwo$ from Ref.~\cite{ALICE:2022zks}.}
\label{fig:prompt_v2_midcentral}
\end{figure}

The left panel of Figure \ref{fig:prompt_v2_midcentral} shows the measured values for the prompt $\vtwo$ of $\Dzero$ and $\Ds$ mesons and $\Lc$ baryons in the $30$–$50\%$ centrality class.
The $\Jpsi$-meson $\vtwo$ from Ref.~\cite{jpsi_v2} and TAMU model~\cite{He:2019vgs} predictions are also reported.
A hint of lower $\vtwo$ for the $\Ds$ meson with respect to the non-strange D mesons can be observed for $1$$\,<\,$$\pt$$\,<\,$$4$ $\GeVc$.
Apart from the small mass difference ($\approx 0.1$ $\GeVc^{2}$), Ref.~\cite{He:2012df} indicates that the different $\vtwo$ at intermediate $\pt$ may stem from hadronic-phase rescattering, which differs between the strange and non-strange cases.
Given the current uncertainties, a firm conclusion is not possible, but the analysis of the data samples collected in 2024 and 2025 will allow a more definitive assessment.
It is evident that the $\Jpsi$-meson $\vtwo$ is lower than the D-meson $\vtwo$ at low and intermediate $\pt$.
This difference is described by the TAMU model, which includes charm-quark interactions with the QGP and hadronization via quark recombination.
By contrast, the difference vanishes for $\pt > 8$ $\GeVc$ where, as seen in Figure \ref{fig:prompt_dmeson_v2_run3_vscentpass4}, also the pion $\vtwo$ converges to similar $\vtwo$ values.
This indicates that the $\vtwo$ originates from the path-length dependence of in-medium energy loss of the partons in the QGP, thus being sensitive to the shape of the generated fireball.
Lastly, the $\Lc$-baryon $\vtwo$ exhibits a splitting with $3.6\sigma$ significance with respect to the $\Dzero$ $\vtwo$ for $\pt > 4$ $\GeVc$, representing the first observation of the baryon–meson $\vtwo$ splitting in the charm sector at the LHC.
The measured $\vtwo$ values of $\Lambda$ and $\text{K}^0_\text{S}$ from Ref.~\cite{ALICE:2022zks} are superimposed with the obtained results for $\Dzero$ and $\Lc$ in the right panel of Figure \ref{fig:prompt_v2_midcentral}.
A clear mass ordering is observed in the low-$\pt$ region while at high $\pt$ the $\vtwo$ values of $\Lambda$ and $\text{K}^0_\text{S}$ respectively group with the ones of $\Lc$ and $\Dzero$.
The observed baryon–meson $\vtwo$ splitting favors a scenario where the anisotropy develops at the quark level, indicating the presence of a deconfined phase in the system evolution.

\begin{figure}[h]
\centering
\includegraphics[width=0.99\textwidth]{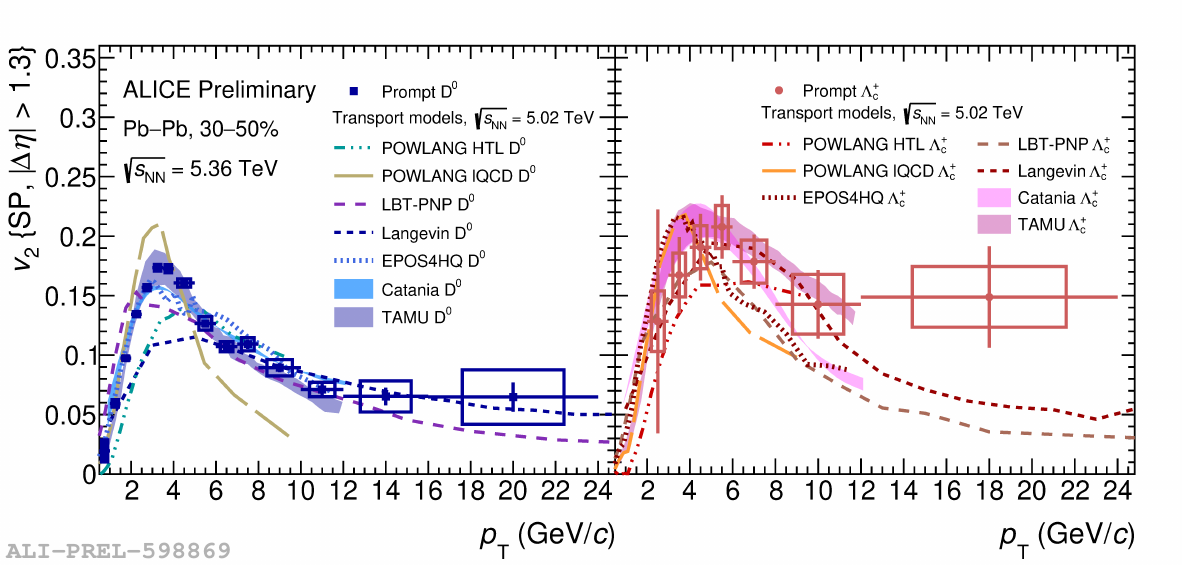}
\caption{Prompt $\Dzero$ and $\Lc$ elliptic flow measurements in the $30$–$50\%$ centrality class superimposed with phenomenological models.}
\label{fig:d0_lc_vs_models}
\end{figure}

In Figure \ref{fig:d0_lc_vs_models}, the measurements of the prompt $\Dzero$ and $\Lc$ $\vtwo$ are superimposed with predictions by state-of-the-art transport models~\cite{He:2019vgs, Li:2020umn, Cassing:2009vt, Sambataro:2024mkr, Beraudo:2023nlq, Xing:2021xwc}.
The models differ among each other in their value for the spatial diffusion coefficient $D_S$, in the hydrodynamic models used for the fireball expansion, in the balance between the coalescence and fragmentation hadronization mechanisms, and in their implementation of the hadronic phase.
The models, which succeed in qualitatively describing the previously published D-meson $\vtwo$ and $\RAA$ measurements within uncertainties, still show tensions among each other and with respect to data.
Therefore, the high-precision measurement of D-meson $\vtwo$ and the newly-released measurement of the $\Lc$ $\vtwo$ can provide powerful constraints to the model parameters.

\begin{figure}[h]
\centering
\includegraphics[width=0.8\textwidth]{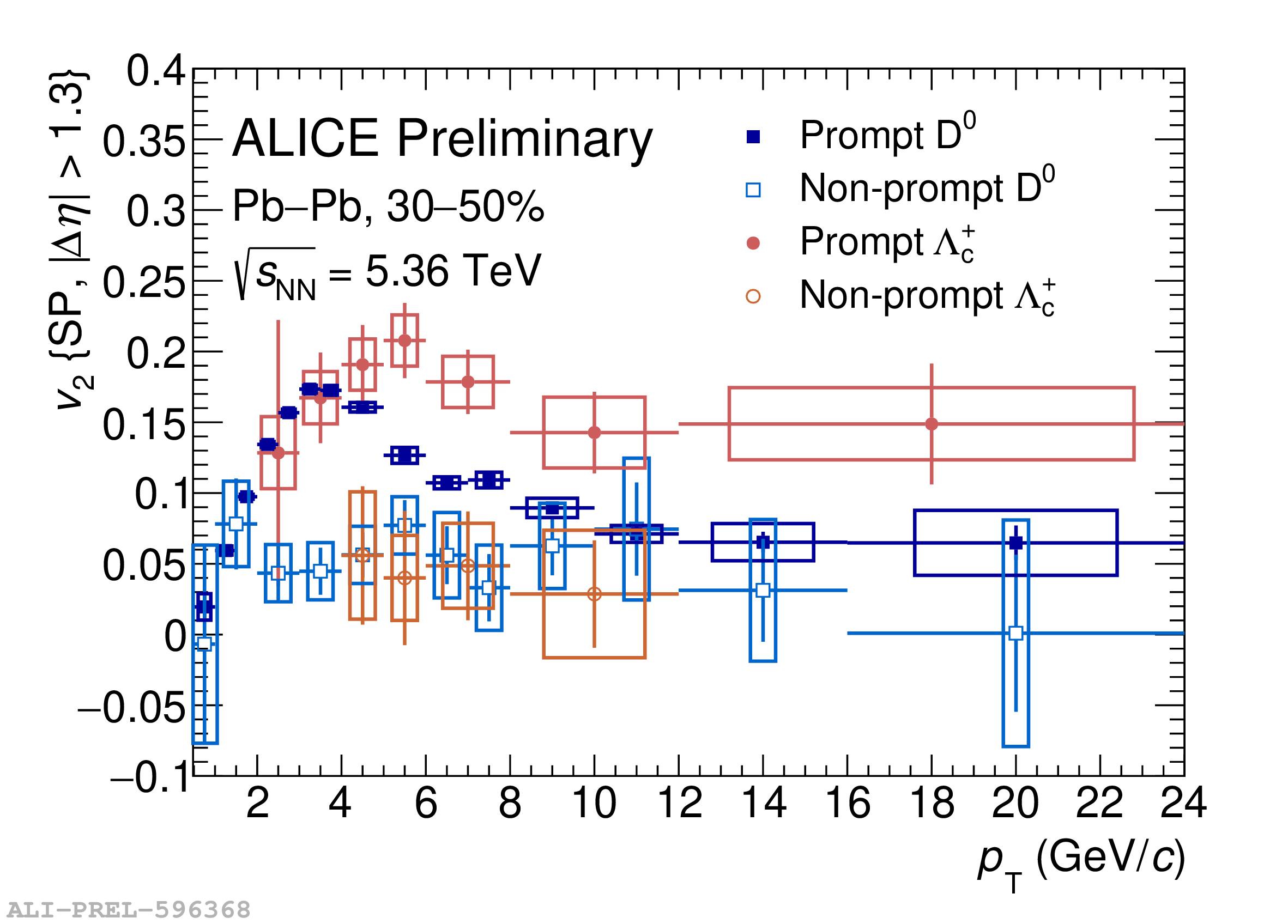}
\caption{Prompt and non-prompt $\Dzero$ and $\Lc$ elliptic flow in semicentral Pb–Pb collisions at $\sqrt{s_{\mathrm{NN}}} = 5.36$ TeV.}
\label{fig:nonprompt_v2_midcentral}
\end{figure}

Figure \ref{fig:nonprompt_v2_midcentral} displays the measurement of the non-prompt $\Dzero$ and $\Lc$ $\vtwo$ in the $30$–$50\%$ centrality class.
The non-prompt hadron $\vtwo$ in the measured $\pt$ interval is positive and lower than that of the corresponding prompt hadrons.
This is expected due to $m_b \gg m_c$, yielding a longer relaxation time in the medium for beauty quarks as compared to charm quarks.
Finally, the $\vtwo$ values of non-prompt $\Dzero$ and $\Lc$ are compatible with each other and no splitting is observed within the current uncertainties.

\section{Conclusion}
In these proceedings, several new results on the study of charm hadron elliptic flow in ultrarelativistic heavy-ion collisions by ALICE using the latest data collected during LHC Run 3 have been presented.
For the first time at the LHC, the prompt and non-prompt $\Lc$-baryon $\vtwo$ is measured in semicentral collisions, providing evidence for the meson–baryon splitting of the $\vtwo$ observable in the charm sector. 
Moreover, the prompt D-meson $\vtwo$ is measured down to 0.5 $\GeVc$ and in peripheral events, where a smaller and shorter-lived QGP phase is produced as compared to central collisions.
A trend of lower $\vtwo$ for the $\Ds$ meson with respect to the non-strange D-meson $\vtwo$ is suggested for $\pt<4$~\GeVc, but the current uncertainties do not support a firm conclusion.
The results will be further improved by including the data sample collected in 2024, similar in size to the one used in this analysis, and the 2025 data, whose projected integrated luminosity is approximately $2.1$ nb$^{-1}$.

\bibliographystyle{utphys}
\bibliography{bibliography}

\end{document}